\newcommand{\diracslash}[1]{#1\llap{/\kern2pt}}

\newcommand{\be}{\begin{equation}}
\newcommand{\ee}{\end{equation}}
\newcommand{\bea}{\begin{eqnarray}}
\newcommand{\eea}{\end{eqnarray}}
\newcommand{\ba}[1]{\begin{array}{#1}}
\newcommand{\ea}{\end{array}}

\newcommand{\bt}{\begin{tabular}}
\newcommand{\et}{\end{tabular}}

\newcommand{\beas}{\begin{eqnarray*}}
\newcommand{\eeas}{\end{eqnarray*}}

\documentclass[preprint,prd,aps,floats,nofootinbib,floatfix]{revtex4}
\usepackage{graphicx}
\begin{document}

\title{D mesons in strongly magnetized asymmetric nuclear matter}
\author{Sushruth Reddy P}
\email{sushruth.p7@gmail.com}
\affiliation{Department of Physics, Indian Institute of Technology, Delhi,
Hauz Khas, New Delhi -- 110 016, India}

\author{Amal Jahan CS}
\email{amaljahan@gmail.com}
\affiliation{Department of Physics, Indian Institute of Technology, Delhi,
Hauz Khas, New Delhi -- 110 016, India}

\author{Nikhil Dhale}
\email{dhalenikhil07@gmail.com}
\affiliation{Department of Physics, Indian Institute of Technology, Delhi,
Hauz Khas, New Delhi -- 110 016, India}

\author{Amruta Mishra}
\email{amruta@physics.iitd.ac.in}
\affiliation{Department of Physics, Indian Institute of Technology, Delhi,
Hauz Khas, New Delhi -- 110 016, India}

\author{Juergen Schaffner-Bielich}
\email{schaffner@astro.uni-frankfurt.de}
\affiliation{Institut fuer Theoretische Physik, 
Goethe Universitaet, Frankfurt,
Max-von-Laue Str. 1,
D-60438,Frankfurt am Main, Germany}

\begin{abstract}
The medium modifications of the open charm mesons ($D$ and $\bar D$)
are studied in isospin asymmetric nuclear matter in the presence
of strong magnetic fields, using a chiral effective model. 
The mass modifications 
of these mesons in the effective hadronic model, arise due to their 
interactions with the protons, neutrons and the scalar mesons 
(non-strange isoscalar $\sigma$, strange isoscalar, $\zeta$
and non-strange isovector, $\delta$), in the magnetized nuclear matter.
In the presence of magnetic field, for the charged baryon, 
i.e., the proton, the number density 
as well as the scalar density have contributions due to the summation 
over the Landau energy levels. For a given value of the baryon
density, $\rho_B$, and isospin asymmetry, 
the scalar fields  
are solved self consistently
from their coupled equations of motion. 
The modifications of the masses of the $D$ and $\bar D$ 
mesons are calculated from the medium modifications
of the scalar fields and the nucleons.
The effects of the anomalous magnetic moments of the nucleons
on the masses of the open charm mesons are also
investigated in the present work. The effects of isospin
asymmetry as well as of the anomalous magnetic moments
are observed to be prominent at high densities for 
large values of magnetic fields.
\end{abstract}
\maketitle

\def\bfm#1{\mbox{\boldmath $#1$}}

\section{Introduction}
The study of the medium modifications of the properties of hadrons 
is an important and challenging area of research in the strong interaction
physics. The topic has attracted a lot of attention in the recent past,
due to its relevance to the ultra-relavitistic heavy ion collision 
experiments. The medium modifications of the hadrons in the matter 
at high densities and/or temperatures resulting 
from these high energy nuclear collisions, affect the 
observables in these experiments. The effects of isospin asymmetry
are also important to investigate, as the heavy ion collision
experiments involve nuclei which have large isospin asymmetry,
with the number of neutrons being much larger than the number of protons.
It is also important to study the effects of magnetic fields 
on the properties of the hadrons in hot and/or dense matter,
since huge magnetic fields are believed to be created 
in the non-central ultra-relativistic heavy ion collision experiments.
The magnetic fields created in the heavy ion experiments
are estimated to be order of $eB \sim 2 m_\pi^2$ (corresponding to a 
magnetic field $\sim 6 \times 10^{18}$ Gauss)
at RHIC, BNL, and, $eB \sim 15 m_\pi^2$ 
at LHC, CERN \cite{kharzeev,skokov}.
Also, strong magnetic fields do exist in astrophysical compact 
objects, like magnetars, which may have magnetic fields of the
order of $10^{15}-10^{16}$ Gauss  \cite{magnetarbb}
at the surface and the magnetic field
could be more intense in the interior of these stars.
The bulk matter existing inside the neutron star 
in the presence of magnetic field
has been studied extensively in the literature 
\cite{somenathprl,broderick1,broderick2,Wei,mao}. 
At high densties, the neutron stars could have a quark core,
comprising of electrically charge neutral strange quark matter
in the presence of magnetic field \cite{somenathprd,ambhasmag}.
The study of strongly interacting matter in magnetic field
has gained a lot of interest in the recent years, due to
the novel effects like chiral magnetic effect \cite{kharzeev}
as well as inverse magnetic catalysis \cite{mag_catalysis}
exhibited in the presence of magnetic field. 

The strong magnetic fields created in non-central ultrarelativistic 
heavy ion collisions decrease rapidly after the collision,
as the ion remnants recede away from the collision zone.
This leads to induced currents
which slow down the decrease in the magnetic field 
\cite{tuchin2011,tuchin2010,Ajit2017}. 
The time evolution of the magnetic field depends crucially on the 
electrical conductivity of the medium,
and the larger the 
electrical conductivity, the longer the magnetic field is
sustained. The effects of the magnetic field in the medium
also tend to increase the electrical conductivity 
\cite{mag_sigma} which slows down the decay of the magnetic field. 
Initially after the collision there is a fast decrease
in the magnetic field, whereas at later times, the 
matter effects lead to slowing 
down the decay of the magnetic field \cite{tuchin2013}. 
The time evolution of magnetic
field in the heavy ion collisions is still an open question. 
This requires proper estimate of the electrical conductivity
of the medium as well as solutions of the magnetohydrodynamic
(MHD) equations, which need further investigations.

The open charm mesons ($D$ and $\bar D$) mesons, have appreciable
modifications in the hadronic medium, because, 
due to the presence of the light quark (antiquark). These
mesons interact with the light quark condensates, which 
are modified significantly in the hadronic medium. 
On the other hand, the charmonium states have modifications
due to interaction with the gluon condensates in the medium,
which are observed to be have much smaller changes
in the medium as compared to those of light quark condensates.
The strong magnetic fields produced in the relativistic heavy
ion collision experiments, e.g. at the RHIC, BNL
and at LHC, CERN, has initiated investigations
of the properties of these heavy flavour mesons,
e.g., $D$ and $B$ mesons, 
\cite{machado_1,machado_2,gubler}, 
as well as charmonium states  \cite{charmonium_mag_lee},
in the presence of magnetic fields. 
The present work is a step in the direction of studying the
masses of the $D$ and $\bar D$ mesons in the presence of 
strong magnetic fields using a chiral effective model.

The open heavy flavour mesons, e.g., $D$ and $\bar D$ mesons,
have been studied in the literature, using the QCD sum rule 
approach \cite{arata,qcdsum08},the Quark Meson Coupling (QMC)
model \cite{qmc_D} (where the quarks within the hadrons
interact via scalar and vector mesons \cite{qmc}),
the effective hadronic model, e.g., chiral SU(4) model
\cite{amdmeson, amarindamprc, amarvdmesonTprc, amarvepja}
as well as using the coupled channel approach 
\cite{ltolos,ljhs,mizutani,HL}.
Due to the presence of a light quark, the scalar open charm
(bottom) are observed to be modified appreciably in the nuclear 
medium \cite{Wang_heavy_scalar}, similar to the $D$ ($B$) meson. 
These heavy flavour mesons
have been studied using QCD sum rule approach
\cite{Wang_heavy_scalar,Wang_heavy_meson,Hilger_scalar_open_charm,
Hilger_sc_PS_open_charm,arvind_heavy_mesons_QSR}
as well as coupled channel approach \cite{tolos_heavy_mesons}.
In the present work, we use a chiral effective model,
for the study of $D$ and $\bar D$ mesons in 
nuclear matter in presence of a magnetic field.
The model based on chiral SU(3) symmetry, 
has been used extensively to study nuclear matter,
finite nuclei \cite{paper3}, hyperonic matter \cite{kristof1}, 
vector mesons \cite{hartree},
kaons and antikaons \cite{kaon_antikaon,isoamss,isoamss1,isoamss2},
as well as to study the charge neutral
matter as the bulk matter comprising the (proto) neutron stars
\cite{pneutronstar}.
The model has been generalized to chiral SU(4), to derive
the interactions of the charm mesons with the light hadronic
sector, so as to investigate the mass modifications of the 
$D$ and $\bar D$ mesons
\cite{amarindamprc, amarvdmesonTprc, amarvepja},
as well as of the strange charm mesons (the $D_s$ mesons)
\cite{DP_AM_Ds}. The model also 
incorporates the broken scale invariance
of QCD \cite{paper3, kristof1,sche1} through a scalar dilaton 
field which mimicks the
gluon condensates of QCD, within the effective hadronic model.
The mass modifications of the heavy quarkonium mesons, 
e.g., charmonium \cite{amarvdmesonTprc,amarvepja} 
and bottomonium states \cite{AM_DP_upsilon}, are due to their
interactions with the gluon condensate of QCD, and, 
within the effective chiral model, these 
arise due to the modifications
of the dilaton field in the hadronic medium. 
The medium modifications of the partial decay widths 
of the charmonium states to $D\bar D$ have been studied
in the literature as arising from the medium modifications
of the $D$ and $\bar D$ meson masses \cite{friman},
using a light quark pair creation model, namely $^3P_0$ model 
\cite{3p0_1,3p0_2}. Using the $^3P_0$ model, 
these decay widths in hadronic matter 
have also been studied, as 
arising due to the mass modifications 
of the open charm mesons and the charmonium states,
calculated within the effective chiral model
\cite{amarvepja}. The in-medium charmonium decay widths
have later been studied, using a field theoretic model
of composite hadrons \cite{amspmwg}. Generalizing the model to the
bottom sector, the in-medium masses of the
$B$, $\bar B$ mesons \cite{AM_DP_bbar} as well as 
the strange bottom meson, $B_s$ mesons \cite{DP_AM_Bs},
as arising from their interactions to the baryons 
and the scalar mesons, have been studied.  
The open charm and open bottom mesons have been studied
in the literature using pion exchange in an effective hadronic
model \cite{sudoh} and the attractive interactions of the
$\bar D$ and $B$ mesons in nuclear matter predict possibility
of the bound states of these mesons with the nuclei \cite{qmc_D}.
Due to attractive interaction of $J/\psi$ in nuclear matter
\cite{leeko,krein_jpsi,amarvdmesonTprc,amarvepja},
the bound states of the $J/\psi$ to nuclei have also been
predicted within the QMC model \cite{krein_17}.
Using the medium modifications of the masses of the
open bottom mesons, $B$ and $\bar B$, as well as the bottomonium
states, as calculated within the effective hadronic model,
the partial decay widths of the $\Upsilon$-states to $B\bar B$,
in hadronic matter have been studied using the field theoretic
model of composite hadrons with quark constituents
\cite{amspm_upsilon}.

As has already been mentioned, the medium modifications
of the masses of the $D$ and $\bar D$ mesons,
arise due to their interactions 
with the baryons and the scalar mesons in isospin 
(a)symmetric hadronic matter. 
The scalar mesons as calculated in the chiral model
in the strange hadronic matter are related to the
light quark condensates, i.e., the non-strange
$\langle \bar q q\rangle$, $q=u,d$, as well as
the strange condensate $\langle \bar s s\rangle$ 
in the medium. From the medium modifications of
the scalar mesons, which are related to the
quark condensates of QCD,  
the in-medium masses of the light vector mesons, namely,
the $\omega$, $\rho$ and $\phi$ mesons \cite{am_vecmeson},
as well as the charmonium states, $J/\psi$ and $\eta_c$
\cite{amarvjpsi_qsr} have been studied, using a
QCD sum rule approach. 
The $D$ and $B$ mesons in the presence of an external magnetic field
have been studied in the literature using a semiclassical 
approach \cite{machado_1}, where the mass of the open charm (bottom)
meson is due to the interaction of the magnetic field
to the spin of the quarks. The magnetic field was observed 
to lead to different masses for the different spin orientations
of the heavy-light quark-antiquark systems.
In the presence of strong magnetic fields, there is mixing 
of the spin 0 states $D$ ($B$) to the spin 1 states $D^*$ ($B^*$).
The D and B mesons were observed to have lowering of their masses
(arising from the mass reductions of specific spin orientations)
\cite{machado_1}. 
The effects of these mass reductions
on the production of the charmonium and bottomonium states were studied 
using the color evaporation model in Ref. \cite{machado_1}. 
The $D$ mesons in the presence of a magnetic field
have also been studied within the QCD sum rule approach,
accounting for the mixing of the pesudoscalar mesons
and the vector mesons, as well as the
Landau quantization effects for the charged $D$ mesons \cite{gubler}.
This showed an increase in the mass of the neutral $D$ mesons,
whereas, the mass spectra of the charged $D$ mesons turned out to saturate,
in the presence of mixing as well as Landau quantization 
effects. The mass of charged $B$ mesons within the QCD sum rule
approach were observed to decrease in the presence of magnetic 
fields \cite{machado_2}.
The mass modifications of the charmonium ground states
($\eta_c$ and $J/\psi$) have also
been studied using the QCD sum rule approach \cite{charmonium_mag_lee}
accounting for the effects of the mixing between these pesudoscalar
and vector mesons.
In the present work, the effects of the magnetic fields 
are considered to study the masses of the $D$ and $\bar D$ mesons
in isospin asymmetric nuclear matter, using the chiral effective model. 
The mass modifications of the $D$ and $\bar D$ mesons 
arise due to their interactions with the nucleons and 
the scalar mesons in the magnetized hadronic matter.
For given values of baryon density, $\rho_B$, 
and isospin asymmetry parameter,
$\eta=(\rho_n-\rho_p)/(2\rho_B)$, 
in the mean field approximation, the values of the scalar meson 
fields,  $\sigma$ ($\sim (u\bar u+d \bar d$)), $\zeta$
($\sim s \bar s$) and $\delta$ ($\sim (u\bar u -d\bar d$)),
are calculated self-consistently from their equations 
of motion. There are contributions from the Landau energy
levels for the number density as well as the scalar density,
for the proton, the charged baryon, in the presence of the
external magnetic field. 
The $D (D^0, D^+)$ and $\bar D (\bar {D^0}, D^-)$ meson
in-medium masses, due to their interactions
with the nucleons as well as the scalar mesons,
are studied 
in the present work. The charged $D$ mesons
($D^{\pm}$ mesons),
additionally have a positive shift in their masses,
due to Landau quantization, in the presence of the 
external magnetic field. 
The effects of anomalous 
magnetic moments of the proton and neutron
\cite{broderick1,broderick2,Wei,mao,amm,VD_SS,aguirre_fermion,aguirre_meson}
are also
investigated and these contributions on the $D$ meson
masses are compared to the case when these effects are not 
taken into account. 
The effects of magnetic fields are
observed to be large at high densities.
At large magnetic fields, the effects
of the anomalous magnetic moments are seen to be significant
at high densities. The effects of isospin asymmetry 
is observed to be large at high densities, 
when the magnetic field is increased. 
In the present work,
the effects on the masses of the
$D$ and $\bar D$ mesons due to the mixing of the pesudoscalar 
and vector mesons \cite{machado_1,machado_2,gubler}
in the presence of magnetic fields, have not been taken 
into consideration, 
as these effects are not within the scope of the mean field approximation
used in the chiral effective model.
Moreover, the effects of temperature on the masses of the $D$ 
and $\bar D$ mesons
were observed to be marginal as compared to the effects of the density,
within the chiral effective model \cite{amarvdmesonTprc}. 
As a first step calculation, in the present work,
the mass modifications of the $D$ and $\bar D$ mesons
in asymmetric nuclear matter have been studied
in the presence of strong magnetic fields,
without accounting for the effects 
from temperature.

We organize the paper as follows: We briefly recapitulate the
$SU(3)$-flavor chiral model adopted for the description of the asymmetric
hadronic matter \cite{isoamss1,isoamss2} in the presence of
an external magnetic field, in Section II. The number density
as well as the scalar density of the charged baryon, 
i.e. the proton, have contributions from the Landau energy levels.
We have additionally taken into account the effects of the
anomalous magnetic moments of the proton and neutron
for the study of the properties of the hadrons.
The medium modifications of the 
$D$ and $\bar D$ meson masses are calculated through their interactions 
with the nucleons and scalar mesons, as has been described
in Section III. 
Section IV discusses the results of the 
present investigation, while we summarise our findings and discuss possible 
outlook in Section V.

\section{ The hadronic chiral $SU(3) \times SU(3)$ model }
The effective hadronic chiral Lagrangian density in the presence of 
magnetic field, used in the present work is given as
\be
{\cal L} = {\cal L}_{kin} + \sum_{ W =X,Y,V,{\cal A},u }{\cal L}_{BW}
          + {\cal L}_{vec} + {\cal L}_0 +
{\cal L}_{scalebreak}+ {\cal L}_{SB}+{\cal L}_{mag},
\label{genlag} \ee 
The above Lagrangian density is based on
a nonlinear realization of SU(3) chiral symmetry 
\cite{weinberg,coleman,bardeen} and broken scale invariance
\cite{paper3,sche1,kristof1}.
In Eq. (\ref{genlag}), 
$ {\cal L}_{kin} $ is the kinetic energy term,
$  {\cal L}_{BW}  $ contains the baryon-meson interactions in
which the baryon-scalar meson interaction terms generate the
baryon masses. $ {\cal L}_{vec} $ describes the dynamical mass
generation of the vector mesons via couplings to the scalar fields
and contains additionally quartic self-interactions of the vector
fields.  ${\cal L}_0 $ contains the meson-meson interaction terms
inducing the spontaneous breaking of chiral symmetry, 
${\cal L}_{scalebreak}$ is a scale invariance breaking logarithmic 
potential, $ {\cal L}_{SB} $ describes the explicit chiral symmetry
breaking. The last term, ${\cal L}_{mag}$ is the contribution 
from the magnetic field, given as
\be 
{\cal L}_{mag}=-{\bar {\psi_i}}q_i 
\gamma_\mu A^\mu \psi_i
-\frac {1}{4} \kappa_i \mu_N {\bar {\psi_i}} \sigma ^{\mu \nu}F_{\mu \nu}
\psi_i
-\frac{1}{4} F^{\mu \nu} F_{\mu \nu}.
\label{lmag}
\ee
In the above, $\psi_i$ corresponds to the $i$-th baryon.
The second term in equation (\ref{lmag}) corresponds 
to the tensorial interaction
with the electromagnetic field and is related to the
anomalous magnetic moments of the baryons (proton and neutron,
in the present investigation). In this term, $\mu_N$ is the Nuclear 
Bohr magneton, given as $\mu_N={e}/({2m_N})$, where 
$m_N$ is the vacuum mass of the nucleon.
We choose the magnetic field to be uniform and along the
z-axis, and take the vector potential to be
$A^\mu =(0,0,Bx,0)$. 

To investigate the hadronic properties in the medium, we write
the Lagrangian density within the chiral SU(3) model in the mean
field approximation and determine the expectation values
of the meson fields by minimizing the thermodynamical potential
\cite{hartree,kristof1}. In the present work, we shall be using 
the frozen glueball
approximation, i.e., fix $\chi=\chi_0$, the vacuum value of
the dilaton field, $\chi$. This is due to the reason that 
the medium modification of the dilaton field is observed
to be negligible as compared to the medium changes of the
scalar fields, $\sigma$, $\zeta$ and $\delta$, and hence
has marginal contribution to the
medium modifications of the $D$ and $\bar D$ meson masses. 
The  non-strange scalar field $\sigma$, 
strange scalar field $ \zeta$ and scalar-isovector field $ \delta$,
are solved from the equations of motion, 
derived from the Lagrangian density. These coupled equations of 
motion are given as
\begin{eqnarray}
&& k_{0}\chi^{2}\sigma-4k_{1}\left( \sigma^{2}+\zeta^{2}
+\delta^{2}\right)\sigma-2k_{2}\left( \sigma^{3}+3\sigma\delta^{2}\right)
-2k_{3}\chi\sigma\zeta \nonumber\\
&-&\frac{d}{3} \chi^{4} \bigg (\frac{2\sigma}{\sigma^{2}-\delta^{2}}\bigg )
+\left( \frac{\chi}{\chi_{0}}\right) ^{2}m_{\pi}^{2}f_{\pi}
-\sum _i  g_{\sigma i}\rho_{i}^{s} = 0 
\label{sigma}
\end{eqnarray}
\begin{eqnarray}
&& k_{0}\chi^{2}\zeta-4k_{1}\left( \sigma^{2}+\zeta^{2}+\delta^{2}\right)
\zeta-4k_{2}\zeta^{3}-k_{3}\chi\left( \sigma^{2}-\delta^{2}\right)\nonumber\\
&-&\frac{d}{3}\frac{\chi^{4}}{\zeta}+\left(\frac{\chi}{\chi_{0}} \right) 
^{2}\left[ \sqrt{2}m_{k}^{2}f_{k}-\frac{1}{\sqrt{2}} m_{\pi}^{2}f_{\pi}\right]
 -\sum_i g_{\zeta i}\rho_{i}^{s} = 0 
\label{zeta}
\end{eqnarray}
\begin{eqnarray}
& & k_{0}\chi^{2}\delta-4k_{1}\left( \sigma^{2}+\zeta^{2}+\delta^{2}\right)
\delta-2k_{2}\left( \delta^{3}+3\sigma^{2}\delta\right) +k_{3}\chi\delta 
\zeta \nonumber\\
& + &  \frac{2}{3} d \left( \frac{\delta}{\sigma^{2}-\delta^{2}}\right)
-\sum_i g_{\delta i}\rho_{i}^{s} = 0
\label{delta}
\end{eqnarray}
In the above, $\rho^s_i$, $i=p,n$, are the scalar densities 
for the proton and neutron respectively, in the magnetic field, $B$ chosen
to be along the z-direction. For proton, the number density as well as
the scalar density have contributions from the 
Landau energy levels in the presence of the magnetic field.
The number density and the scalar density of the proton are given as
\cite{Wei,mao}

\begin{equation}
\rho_p=\frac{eB}{4\pi^2} \Bigg [ 
\sum_{\nu=0}^{\nu_{max}^{(S=1)}} k_{f,\nu,1}^{(p)} 
+\sum_{\nu=1}^{\nu_{(max)}^{(S=-1)}} k_{f,\nu,-1}^{(p)} 
\Bigg]
\end{equation}
and 
\begin{eqnarray}
\rho^s_p & = & \frac{eB{m_p^*}}{2\pi^2} \Bigg [ 
\sum_{\nu=0}^{\nu_{max}^{(S=1)}}
\frac {\sqrt {{m_p^*}^2+2eB\nu}+\Delta_p}{\sqrt {{m_p^*}^2+2eB\nu}}
\ln |\frac{ k_{f,\nu,1}^{(p)} + E_f^{(p)}}{\sqrt {{m_p^*}^2
+2eB\nu}+\Delta_p}|\nonumber \\
 &+&\sum_{\nu=1}^{\nu_{max}^{(S=-1)}}
\frac {\sqrt {{m_p^*}^2+2eB\nu}-\Delta_p}{\sqrt {{m_p^*}^2+2eB\nu}}
\ln |\frac{ k_{f,\nu,-1}^{(p)} + E_f^{(p)}}{\sqrt {{m_p^*}^2
+2eB\nu}-\Delta_p}|\Bigg ]
\end{eqnarray}
where, $k_{f,\nu,\pm 1}^{(p)}$ are the Fermi momenta of protons
for the Landau level, $\nu$ for the spin index, $S=\pm 1$,
i.e. for spin up and spin down projections for the proton.
These Fermi momenta are related to the Fermi energy of the
proton as
\begin{equation}
k_{f,\nu,S}^{(p)}=\sqrt { {E_f^{(p)}}^2
-\Big (
{\sqrt {{m_p^*}^2+2eB\nu}+S\Delta_p}\Big )^2}.
\end{equation}
The number density and the scalar density of neutrons are given as
\begin{equation}
\rho_{n}= \frac{1}{4\pi^2} \sum _{S=\pm 1}
\Bigg \{ \frac{2}{3} {k_{f,S}^{(n)}}^3
+S\Delta_n \Bigg[ (m_n^*+S\Delta_n) k_{f,S}^{(n)}
+{E_f^{(n)}}^2 \Bigg( arcsin \Big (
\frac{m_n^*+S\Delta_n}{E_f^{(n)}}\Big)-\frac{\pi}{2}\Bigg)\Bigg]
\Bigg \}
\end{equation}
and
\begin{equation}
\rho^s_n =\frac{m_n^*}{4\pi^2} \sum _{S=\pm 1} 
\Bigg [ k_{f,S}^{(n)} E_f^{(n)} - 
(m_n^*+S\Delta_n)^2 \ln | \frac {k_{f,S}^{(n)}+ 
E_f^{(n)}}{m_n^*+S\Delta_n} | \Bigg].
\end{equation}
The Fermi momentum, $k_{f,S}^{(n)}$ 
for the neutron with spin projection, S 
($S=\pm 1$ for the up (down) spin projection), 
is related to the Fermi energy for the 
neutron, $E_f^{(n)}$ as
\begin{equation}
k_{f,S}^{(n)}= \sqrt { {E_f^{(n)}}^2 -
(m_n^*+S\Delta_n)^2}.
\end{equation}
In the above, the parameter $\Delta _{i}$ refers 
to the anomalous magnetic moment for the baryon, $i$
($i=p,n$), given as
\begin{equation}
\Delta_i =-\frac{1}{2} \kappa_i \mu_N B,
\end{equation}
where, $\kappa_i$,
is as defined in the electromagnetic tensor term
in the Lagrangian density given by (\ref{lmag}).
The values of $\kappa_p$ and  $\kappa_n$
are given as 
$3.5856$ and $-3.8263$ respectively, which are the values
of the gyromagnetic ratio corresponding to the 
anomalous magnetic moments of the proton and 
neutron respectively.

In the following section, we shall study the medium modifications of the
$D$ and $\bar D$ mesons in the isopsin asymmetric nuclear matter
in the presence of a magnetic field. We shall consider the effects of
the anomalous magnetic moments of the baryons,
on the mass modifications of these mesons, and compare the
results for the $D$ and $\bar D$ meson masses,
when these effects are not taken into account.

\section{In medium masses of $D$ and $\bar D$ mesons in magnetized
isospin asymmetric nuclear matter}

The in-medium changes of the $D$ and $\bar D$ mesons in asymmetric
nuclear matter are studied in the presence of strong external 
magnetic fields. These medium modifications are investigated
using a chiral effective model, where the chiral SU(3) 
has been generalized to chiral SU(4) in order to derive the
interactions of the charmed mesons with the light hadronic sector
\cite{amdmeson,amarindamprc,amarvdmesonTprc,amarvepja}.

The interaction Lagrangian modifying the $D$-meson mass can be written
as \cite{isoamss1}
\begin{eqnarray}
\cal L _{DN} & = & -\frac {i}{8 f_D^2} \Big [3\Big (\bar p \gamma^\mu p
+\bar n \gamma ^\mu n \Big) 
\Big({D^0} (\partial_\mu \bar D^0) - (\partial_\mu {{D^0}}) {\bar D}^0 \Big )
+\Big(D^+ (\partial_\mu D^-) - (\partial_\mu {D^+})  D^- \Big )
\nonumber \\
& +&
\Big (\bar p \gamma^\mu p -\bar n \gamma ^\mu n \Big) 
\Big({D^0} (\partial_\mu \bar D^0) - (\partial_\mu {{D^0}}) {\bar D}^0 \Big )
- \Big( D^+ (\partial_\mu D^-) - (\partial_\mu {D^+})  D^- \Big )
\Big ]
\nonumber \\
 &+ & \frac{m_D^2}{2f_D} \Big [ 
(\sigma +\sqrt 2 \zeta_c)\big (\bar D^0 { D^0}+(D^- D^+) \big )
 +\delta \big (\bar D^0 { D^0})-(D^- D^+) \big )
\Big ] \nonumber \\
& - & \frac {1}{f_D}\Big [ 
(\sigma +\sqrt 2 \zeta_c )
\Big ((\partial _\mu {{\bar D}^0})(\partial ^\mu {D^0})
+(\partial _\mu {D^-})(\partial ^\mu {D^+}) \Big )
\nonumber \\
 & + & \delta
\Big ((\partial _\mu {{\bar D}^0})(\partial ^\mu {D^0})
-(\partial _\mu {D^-})(\partial ^\mu {D^+}) \Big )
\Big ]
\nonumber \\
&+ & \frac {d_1}{2 f_D^2}(\bar p p +\bar n n 
 )\big ( (\partial _\mu {D^-})(\partial ^\mu {D^+})
+(\partial _\mu {{\bar D}^0})(\partial ^\mu {D^0})
\big )
\nonumber \\
&+& \frac {d_2}{4 f_D^2} \Big [
(\bar p p+\bar n n))\big ( 
(\partial_\mu {\bar D}^0)(\partial^\mu {D^0})
+ (\partial_\mu D^-)(\partial^\mu D^+) \big )\nonumber \\
 &+&  (\bar p p -\bar n n) \big ( 
(\partial_\mu {\bar D}^0)(\partial^\mu {D^0})\big )
- (\partial_\mu D^-)(\partial^\mu D^+) ) 
\Big ]
\label{lagd}
\end{eqnarray}
In (\ref{lagd}), the first term is the vectorial Weinberg Tomozawa
interaction term, which is attractive for $D$ mesons,
but repulsive for the $\bar D$ mesons. 
The second term is the scalar meson exchange
term, which is attractive for both $D$ and $\bar D$ mesons.
The third, fourth and fifth terms comprise
the range term in the chiral model. 
The parameters $d_1$ and $d_2$ in the last two terms of the
interaction Lagrangian given by (\ref{lagd}) are determined
by fitting to the empirical values of the KN scattering lengths
\cite{thorsson,juergen,barnes}
for I=0 and I=1 channels \cite{isoamss1,isoamss2}.

The dispersion relations for the $D$ and $\bar D$ mesons 
are obtained from the Fourier transformations of the
equations of motion of these mesons. These are given as
\begin{equation}
-\omega^2+ {\vec k}^2 + m_{D(\bar D)}^2
 -\Pi_{D(\bar D)}(\omega, |\vec k|)=0,
\label{dispddbar}
\end{equation}
where $\Pi_{D(\bar D)}$ denotes the self energy 
of the $D$ ($\bar D$) meson in the medium.
For the $D$ meson doublet ($D^0$,$D^+$), and $\bar D$ meson
doublet (${\bar D}^0$,$D^-$), the self enrgies are given by
\begin{eqnarray}
\Pi (\omega, |\vec k|) &= & \frac {1}{4 f_D^2}\Big [3 (\rho_p +\rho_n)
\pm (\rho_p -\rho_n) \big)
\Big ] \omega \nonumber \\
&+&\frac {m_D^2}{2 f_D} (\sigma ' +\sqrt 2 {\zeta_c} ' \pm \delta ')
\nonumber \\ & +& \Big [- \frac {1}{f_D}
(\sigma ' +\sqrt 2 {\zeta_c} ' \pm \delta ')
+\frac {d_1}{2 f_D ^2} (\rho^s_p +\rho^s_n)\nonumber \\
&+&\frac {d_2}{4 f_D ^2} \Big (({\rho^s_p} +{\rho^s_n})
\pm   ({\rho^s_p} -{\rho^s_n}) \Big ) \Big ]
(\omega ^2 - {\vec k}^2),
\label{selfd}
\end{eqnarray}
and
\begin{eqnarray}
\Pi (\omega, |\vec k|) &= & -\frac {1}{4 f_D^2}\Big [3 (\rho_p +\rho_n)
\pm (\rho_p -\rho_n) \Big ] \omega\nonumber \\
&+&\frac {m_D^2}{2 f_D} (\sigma ' +\sqrt 2 {\zeta_c} ' \pm \delta ')
\nonumber \\ & +& \Big [- \frac {1}{f_D}
(\sigma ' +\sqrt 2 {\zeta_c} ' \pm \delta ')
+\frac {d_1}{2 f_D ^2} (\rho^s_p +\rho^s_n
)\nonumber \\
&+&\frac {d_2}{4 f_D ^2} \Big (({\rho^s_p} +{\rho^s _n})
\pm   ({\rho^s_n} -{\rho^s_n}) \Big ]
(\omega ^2 - {\vec k}^2),
\label{selfdbar}
\end{eqnarray}
where the $\pm$ signs refer to the $D^0$ and $D^+$ respectively
in equation (\ref{selfd}) and 
to the $\bar {D^0}$ and $D^-$ respectively in equation (\ref{selfdbar}).
In equations (\ref{selfd}) and (\ref{selfdbar}), 
$\sigma'(=(\sigma-\sigma _0))$,
${\zeta_c}'(=(\zeta_c-{\zeta_c}_0))$ and  $\delta'(=(\delta-\delta_0))$
are the fluctuations of the scalar-isoscalar fields $\sigma$ and $\zeta_c$,
and the third component of the scalar-isovector field, $\delta$,
from their vacuum expectation values.
The vacuum expectation value of $\delta$ is zero ($\delta_0$=0), since
a nonzero value for it will break the isospin symmetry of the vacuum.
In the above, $\rho_p$ and $\rho_n$ are the number densities
of proton and neutron and $\rho^s_p$ and $\rho^s_n$ are
their scalar densities.

The masses of the charged open charm mesons, $D^{\pm}$ have 
an additional positive mass shift due to the presence of the
magnetic field, which retaining only the lowest Landau level,
is given as
\begin{equation}
m^{eff}_{D^\pm}=\sqrt {{m^*_{D^\pm}}^2 +|eB|},
\label{mdpm_landau}
\end{equation}
whereas for the neutral $D(\bar D)$ mesons, namely,
$D^0 (\bar {D^0})$, the effective masses are given as

\begin{equation}
m^{eff}_{D^0 (\bar {D^0})}=m^*_{D^0 (\bar {D^0})},
\end{equation}
In the above, $m^*_{D(\bar D)}$ are the solutions for $\omega$ 
at $|\vec k|=0$,
of the dispersion relations given by equation 
(\ref{dispddbar}).

In the next section, we shall discuss the results for 
the $D(\bar D)$-meson mass modifications in (a)symmetric nuclear
matter in the presence of an external magnetic field, as
obtained in the present effective chiral model.
\section{Results and Discussions}
\label{results}

The in-medium masses of the $D$ and $\bar D$ mesons are 
investigated in asymmetric nuclear matter in the presence of an external
magnetic field, using a chiral effective model. The medium modifications
arise due to the interactions of these open charm mesons with the
protons, neutrons and the scalar mesons ($\sigma$, $\zeta$ and $\delta$). 
The number density and scalar density of the charged baryon,
the proton, have contributions from the Landau energy levels,
in the presence of the external magnetic field, $\vec B=(0,0,B)$.
The in-medium masses of the $D$ and $\bar D$ mesons
are calculated by using the dispersion 
relation given by (\ref{dispddbar}), with the self-energies for the
$D(D^0,D^+)$ and $\bar D(\bar {D^0},D^-)$ mesons given 
by equations (\ref{selfd}) and (\ref{selfdbar}) respectively. 
The charged 
$D$ and $\bar D$ mesons, i.e., $D^{\pm}$ have additional
mass modifications due to the Landau quantization effects in
the presence of the external magnetic field, as given by
equation (\ref{mdpm_landau}). 

\begin{figure}
\includegraphics[width=16cm,height=16cm]{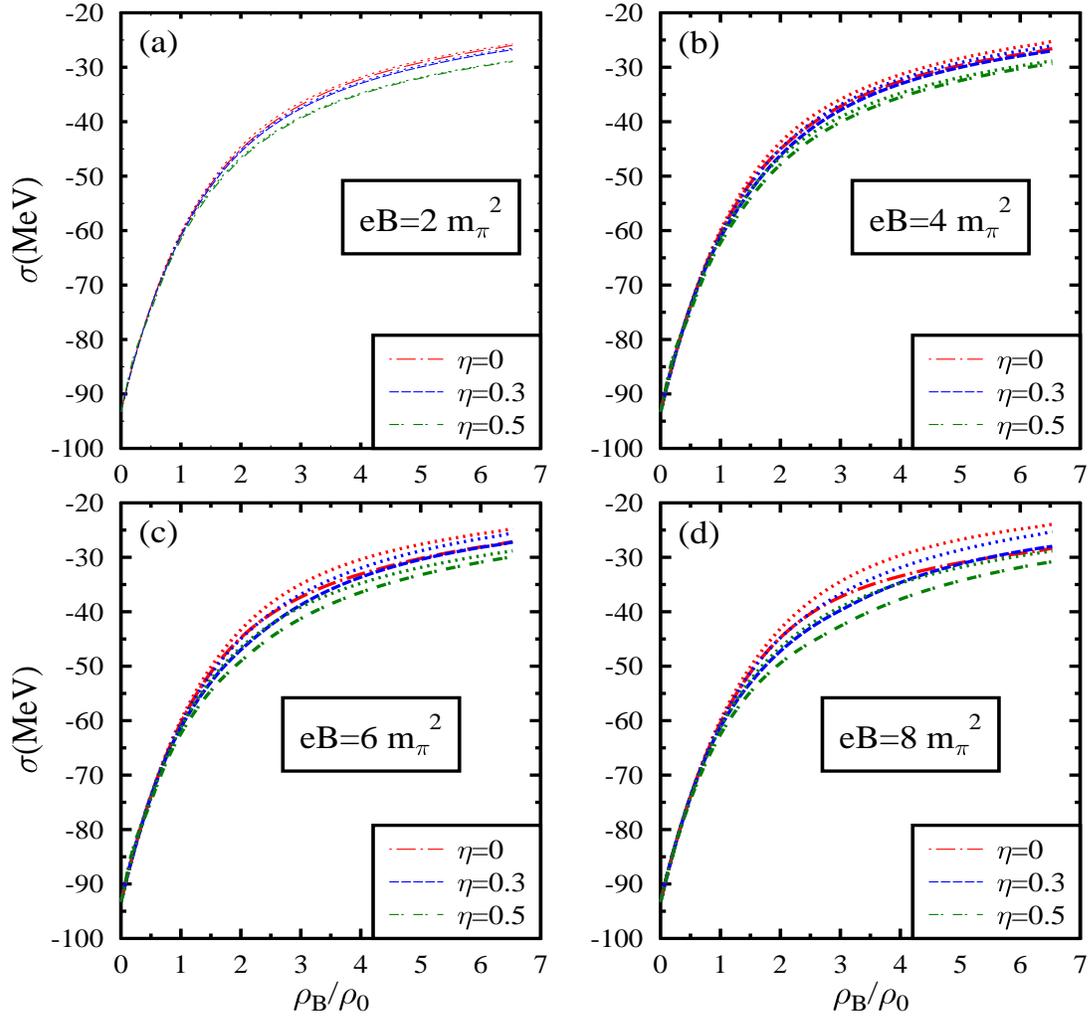}
\caption{
The scalar field $\sigma$ plotted as 
function of the baryon density, $\rho_B/\rho_0$,
for different values of magnetic field, 
is shown for different values of the isospin asymmetry 
parameter, $\eta$, 
when the effects of anomalous magnetic moment are
taken into account, and are compared to the case
when the effects of anomalous magnetic moment is not taken
into account (dotted line).
\label{sigma_mag}
}
\end{figure}

\begin{figure}
\includegraphics[width=16cm,height=16cm]{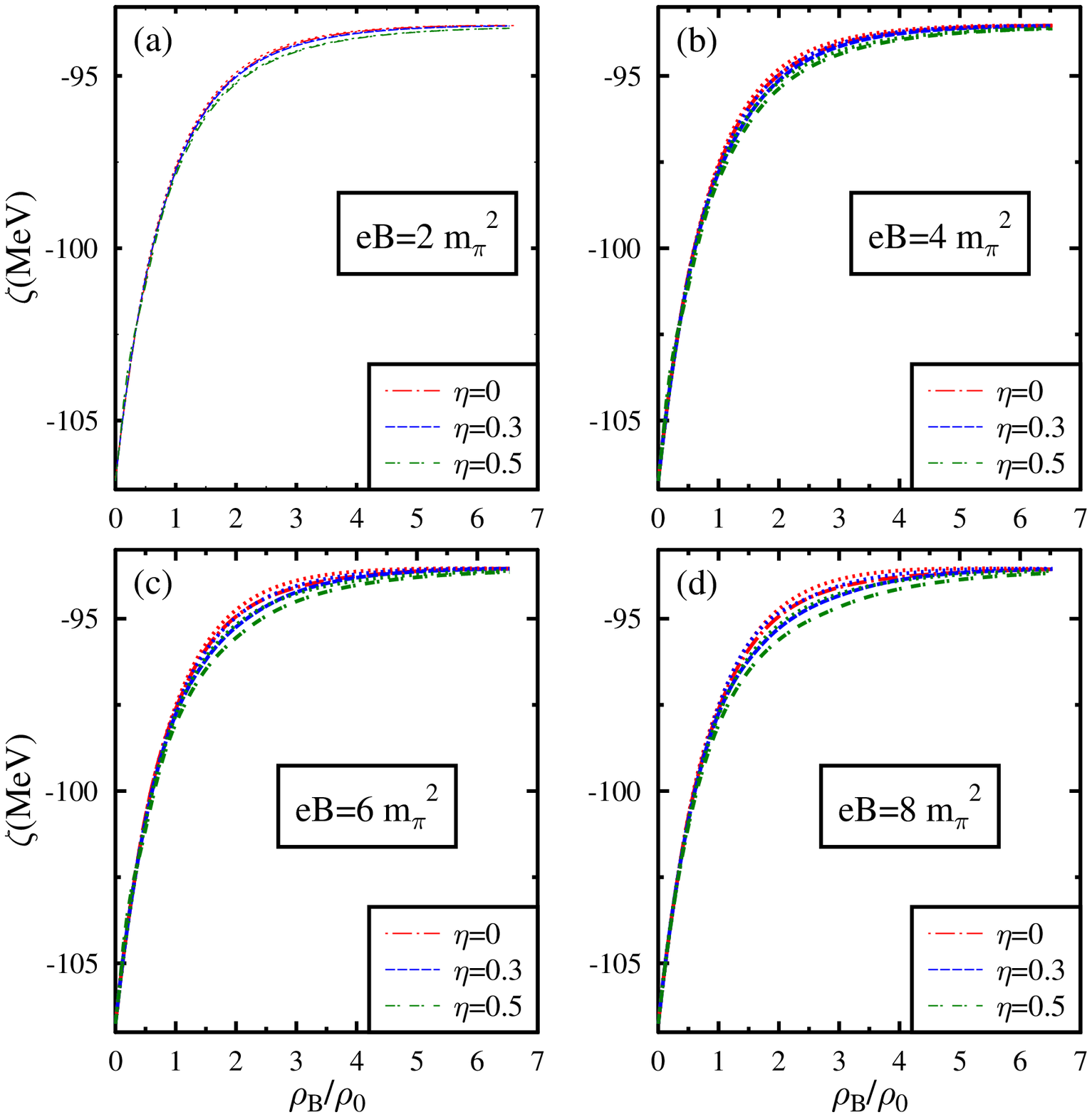}
\caption{
The scalar field $\zeta$ plotted as 
function of the baryon density, $\rho_B/\rho_0$,
for different values of magnetic field, 
is shown for different values of the isospin asymmetry 
parameter, $\eta$, 
when the effects of anomalous magnetic moment are
taken into account, and are compared to the case
when the effects of anomalous magnetic moment is not taken
into account (dotted line).
\label{zeta_mag}
}
\end{figure}

\begin{figure}
\includegraphics[width=16cm,height=16cm]{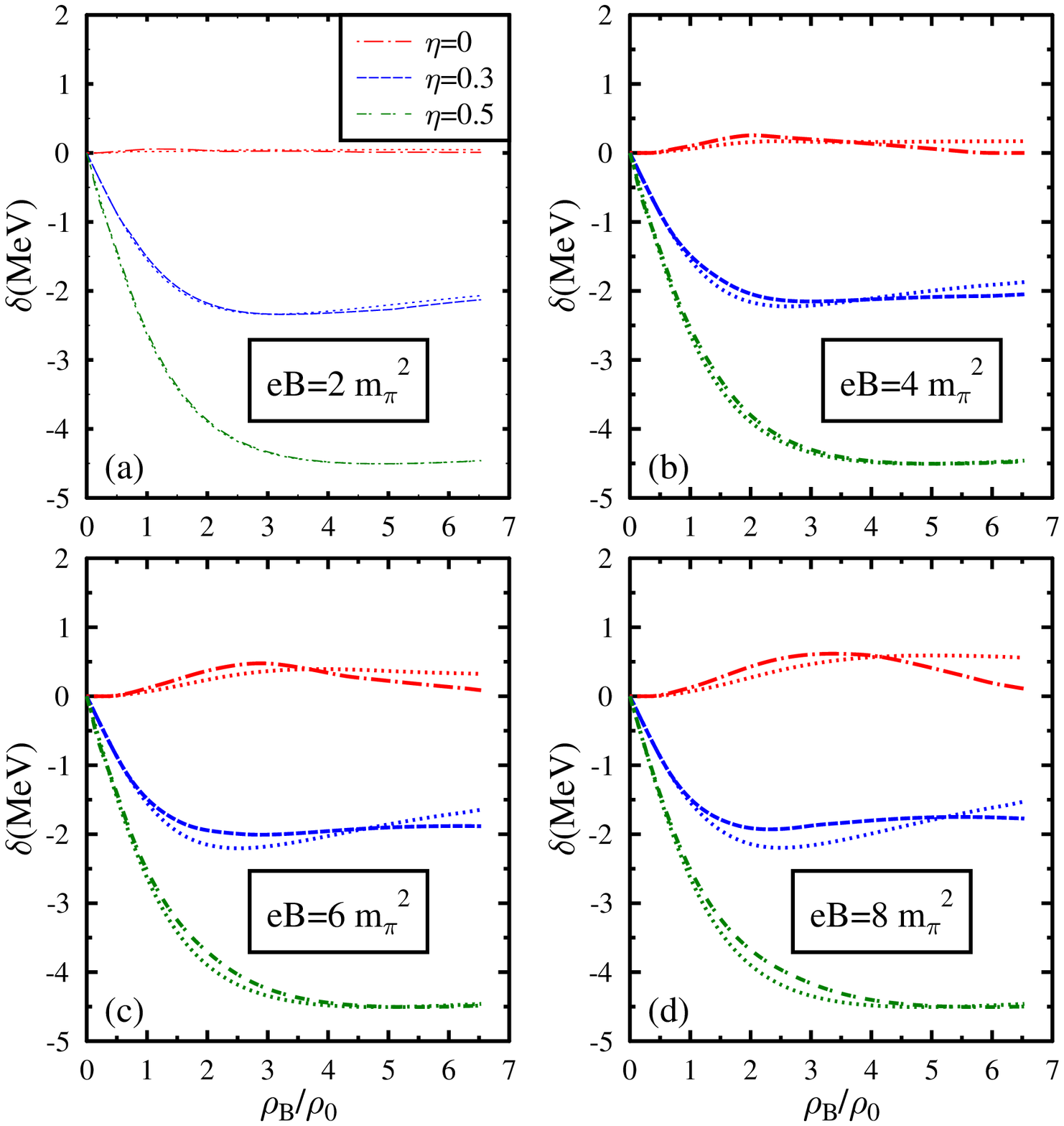}
\caption{
The scalar field $\delta$ plotted as 
function of the baryon density, $\rho_B/\rho_0$,
for different values of magnetic field, 
is shown for different values of the isospin asymmetry 
parameter, $\eta$, 
when the effects of anomalous magnetic moment are
taken into account, and are compared to the case
when the effects of anomalous magnetic moment is not taken
into account (dotted line).
\label{delta_mag}
}
\end{figure}

The expectation values of the scalar fields, $\sigma$,
$\zeta$ and $\delta$, are solved from the equations of motion 
of these mesons, given by equations (\ref{sigma}), (\ref{zeta})
and (\ref{delta}). The values of $\sigma$, $\zeta$ and $\delta$
are plotted as functions of $\rho_B/\rho_0$ 
(the baryon density in units of the nuclear matter saturation density) 
in figures \ref{sigma_mag}, \ref{zeta_mag}, 
and \ref{delta_mag}, respectively.
These are shown for various values of the magnetic field,
as well as for different values of the isospin asymmetry parameter, 
$\eta=(\rho_n-\rho_p)/(2 \rho_B)$, including the effects from
the anomalous magnetic moments (AMM) of the nucleons. These are compared
with the case of without accounting for the effects from
the anomalous magnetic moments (shown as dotted lines).
In figure \ref{sigma_mag}, the isospin asymmetry effects
on the scalar field $\sigma$ are illustrated for values of
the magnetic field $eB$ as $2m_\pi ^2$,  $4m_\pi ^2$,  
$6m_\pi ^2$, and  $8m_\pi ^2$, in panels (a), (b), (c) and
(d) respectively. For isospin symmetric nuclear matter, the effect of 
the anomalous magnetic moments is observed to give 
smaller values of the scalar densities  of the proton and neutron,
as compared to the case when these effects are neglected. This leads
to the magnitude of the $\sigma$ with AMM effects, 
to be larger, as can be inferred
from the equation for $\sigma$ given by equation (\ref{sigma}).  
There is observed to be an increase
in the magnitude of the value of $\sigma$, when one
goes from symmetric nuclear matter to asymmetric nuclear matter.
This is a reflection of the fact that the value of $(\rho_p^s+\rho_n^s)$ 
is smaller for the isospin asymmetric case.
The effects of the isospin asymetry as well as anomalous magnetic
moments are seen to be larger at high densities.
These effects are observed to be larger with increase 
in the magnetic field, as can be seen in figure \ref{sigma_mag}.  
For the case of $\zeta$ meson,
the behaviour of the expectation value of the field, 
is seen to be similar to that of the $\sigma$
meson. However, the change is observed to be much
smaller than that of the $\sigma$ field.

In figure \ref {delta_mag}, the $\delta$ field is plotted 
as a function of the baryon density
(in units of the nuclear matter saturation density,
$\rho_0$), for different values of magnetic fields,
and the effect of the isospin asymmetry on the value 
of $\delta$ is also investigated. For the isospin symmetric 
case ($\eta$=0, i.e., $\rho_p=\rho_n$), the value of $\delta$
(which is determined by the difference of $(\rho^s_p-\rho_s^n)$),
is zero for the case when there is no external magnetic  
field, as $\rho_p^s=\rho_n^s$ in this case. This is because
the Fermi momentum for the proton is equal to that of the
neutron for $\rho_p=\rho_n$, and hence $\rho_p^s=\rho_n^s$, 
for the case of zero magnetic field, which gives the value of
$\delta$ to be zero for symmetric nuclear matter
($\eta$=0) in the absence of magnetic field. 
However, in the presence of magnetic field, 
in symmetric nuclear matter, 
the proton, being the charged nucleon, has contributions 
from the Landau levels. The scalar density of proton, $\rho_p^s$ 
thus turns out to be different from the 
scalar density of neutron,
$\rho_n^s$, and the value of $\delta$, as obtained from the solutions
of the coupled equations for the scalar fields, given by
equations (\ref{sigma}), (\ref{zeta}) and (\ref{delta}),
is found to be nonzero and positive,
as can be seen from figure \ref{delta_mag}. 
When the anomalous magnetic moments (AMM) of the nucleons  
are taken into account, for symmetric nuclear matter ($\eta$=0)
there is observed to be an intial increase in the value of 
$\delta$ with density reaching a maximum value, 
followed by a drop when the density is further increased.
The value of $\delta$ is observed to increase with increase 
in the magnetic field,
reaching a maximum value of around 0.6 MeV, for $eB=8 m_\pi^2$
at a density of around $3\rho_0$, when the effects of anomalous magnetic
moments (AMM) of the nucleons are taken into account. The values of $\delta$
for $\eta$=0 remain similar when the AMM effects are not taken 
into account. However, there is observed to be a monotonic increase
in the value of $\delta$ with density in the latter case. 
For isospin asymmetry parameter, $\eta$=0.3,
in the absence of AMM effects,
there is observed to be an initial increase in the magnitude of 
the $\delta$ field upto a density of around 2$\rho_0$,
followed by a drop in the magnitude when the density is further
increased. The behaviour is related to the similar behaviour of 
$(\rho_p^s-\rho_n^s)$, with density, for this value of the
isospin asymmetry parameter ($\eta$=0.3).
Inclusion of the effects of anomalous magnetic moments, leads to
first a drop, i.e., $\delta$ becomes more negative,
and then a slight increase in the value of $\delta$
with density, for $eB=2 m_\pi^2$, whereas for the higher values 
of the magnetic field, the value remains almost unchanged at densities
higher than around 2.5$\rho_0$. For the maximum asymmetric case
($\eta$=0.5), i.e., when there are only neutrons in the system, 
the anomalous magnetic moment due to the neutron
is observed to give rise to a smaller magnitude
for the $\delta$ field, for densities in the range of
$\rho_0$ to 4$\rho_0$. For densities larger than $4\rho_0$,
the value of $\delta$ is observed to be very similar for 
both the cases of with and without accounting for the anomalous 
magnetic moment effects, for $\eta$=0.5.

The dependence of the scalar fields, $\sigma$, $\zeta$
and $\delta$ on the baryon density is observed to be the 
dominant medium effect as compared to the dependence of these
fields on the magnetic field, as can be seen from figures
\ref{sigma_mag}, \ref{zeta_mag} and \ref{delta_mag}.
In isospin symmetric matter, 
at $\rho_B=\rho_0$,  the value of $\sigma$ 
changes only marginally
with increase in magnetic field. For example, the values 
of $\sigma$ (in MeV)
at $eB$ as equal to $m_\pi^2$, $5m_\pi^2$ and $8m_\pi^2$, 
are observed to be $-60.54 (-60.31)$, $-60.4 (-59.94)$ 
and $-60.42 (-59.9)$ respectively, with (without) AMM effects.
For $\eta$=0.3, these values of $\sigma$ (in MeV) are modified to
$-60.83 (-60.71)$, $-61.45 (-60.61)$ and $-61.58(-60.62)$ respectively.
As the density is increased to $5\rho_0$, the value of 
$\sigma$ is observed to be $-29 (-28.91)$, $-29.835 (-28.07)$
and $-31 (-26.74)$ for symmetric nuclear matter 
with (without) accounting for the AMM effects,
for the value of $eB$ as equal to $m_\pi^2$, $5m_\pi^2$ and
$8m_\pi^2$ respectively. 
The scalar field $\zeta$ follows the same trend as $\sigma$,
but the changes in case of $\zeta$ are found to be extremely small 
even at higher densities and high magnetic fields. 
The value of $\delta$ at small 
densities and low magnetic field is close to zero 
in symmetric nuclear matter. With the increase 
in density the value of $\delta$ becomes more noticeable and 
it increases further at higher magnetic fields. 
For $\eta$ = 0.5, and in the absence of the AMM effects,
the scalar fields, $\sigma$, $\zeta$ and $\delta$, 
do not respond to changes 
in magnetic field and remain constant for a
particular density as there are only neutrons in the medium
for $\eta$=0.5, and neutrons are not subjected 
to magnetic field effects, when the anomalous magnetic moment
effects are not taken into consideration.

\begin{figure}
\includegraphics[width=16cm,height=16cm]{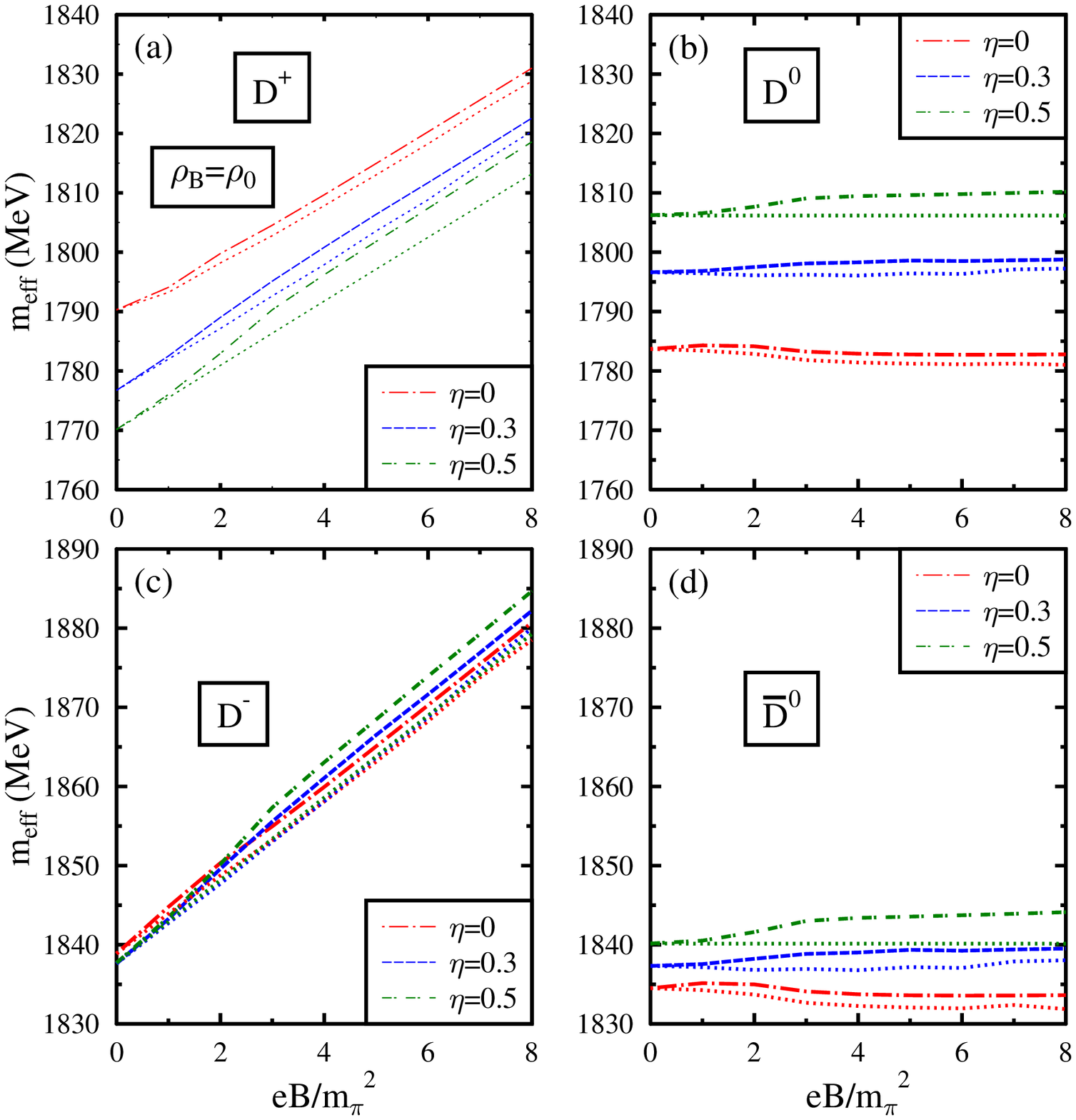}
\caption{
The effective masses of $D$ ($D^+$,$D^0$) and $\bar D$ 
($D^-$, $\bar {D^0}$) mesons 
in MeV plotted as functions of 
$eB/{m_\pi^2}$, 
for baryon density, $\rho_B=\rho_0$,
with different values of  isospin asymmetry parameter, $\eta$,
accounting for the effects of
the anomalous magnetic moments (AMM) for the nucleons.
The results are compared with the case of not accounting for the
anomalous magnetic moments (shown as dotted lines).
\label{mddbar_mag_rhb0}
}
\end{figure}

\begin{figure}
\includegraphics[width=16cm,height=16cm]{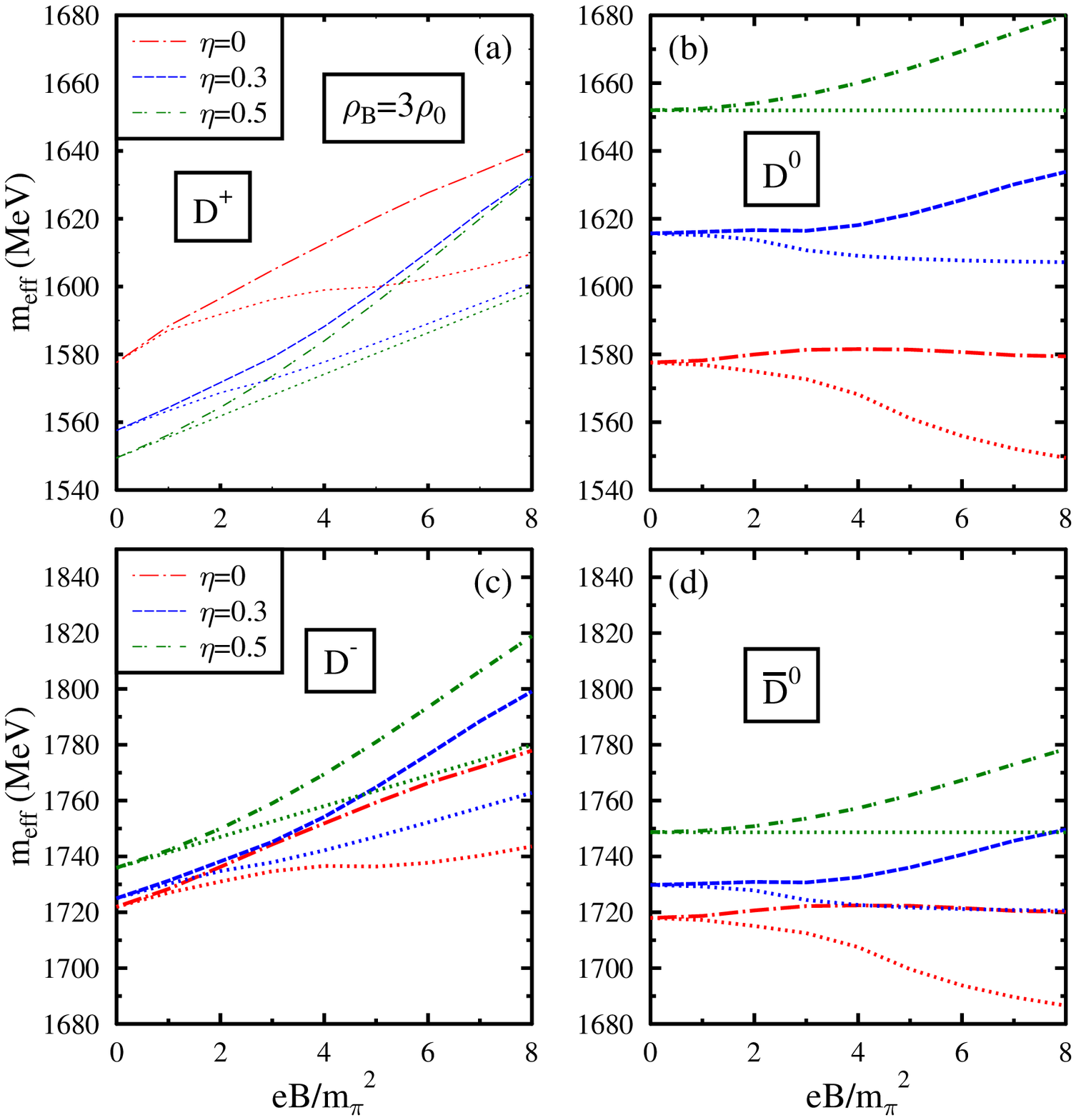}
\caption{
The effective masses of $D$ ($D^+$,$D^0$) and $\bar D$ 
($D^-$, $\bar {D^0}$) mesons in MeV plotted as functions of 
$eB/{m_\pi^2}$, for baryon density, $\rho_B=3\rho_0$,
with different values of  isospin asymmetry parameter, $\eta$,
accounting for the effects of
the anomalous magnetic moments (AMM) for the nucleons.
The results are compared with the case of not accounting for the
anomalous magnetic moments (shown as dotted lines).
\label{mddbar_mag_3rhb0}
}
\end{figure}

\begin{figure}
\includegraphics[width=16cm,height=16cm]{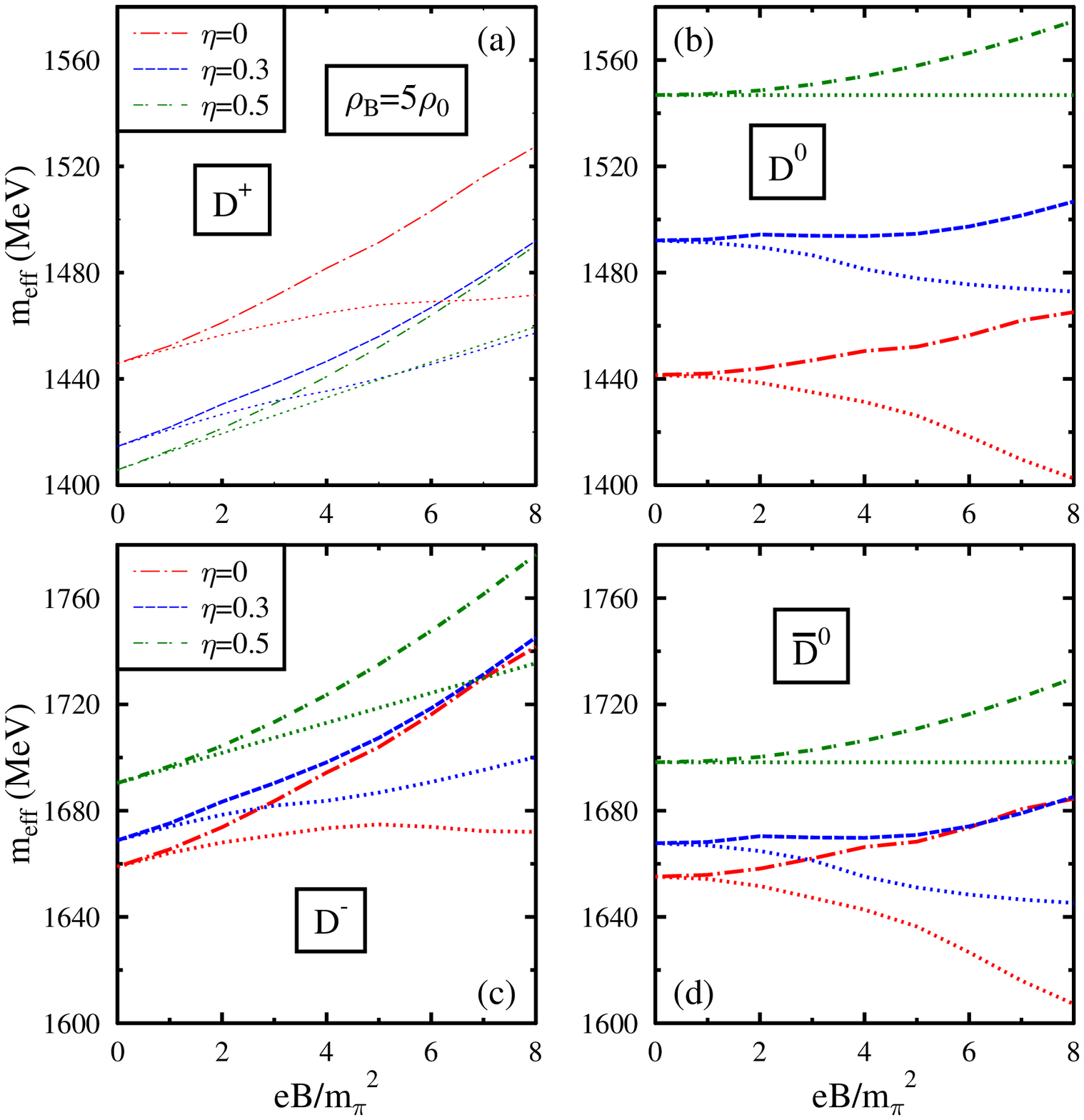}
\caption{
The effective masses of $D$ ($D^+$,$D^0$) and $\bar D$ 
($D^-$, $\bar {D^0}$) mesons 
in MeV plotted as functions of 
$eB/{m_\pi^2}$, for baryon density, $\rho_B=5\rho_0$,
with different values of isospin asymmetry parameter, $\eta$,
accounting for the effects of
the anomalous magnetic moments (AMM) for the nucleons.
The results are compared with the case of not accounting for the
anomalous magnetic moments (shown as dotted lines).
\label{mddbar_mag_5rhb0}
}
\end{figure}

The in-medium masses of the $D$ mesons ($D^+$ and $D^0$)
as well as $\bar D$ ($D^-$ and $\bar {D^0}$)
mesons in magnetized nuclear matter are calculated 
from the interactions of these mesons
as given by equation (\ref{lagd}) and, from the Landau
quantization for the charged $D^\pm$ mesons. 
The dependence of the masses of the $D^+$, $D^0$, $D^-$ 
and $\bar {D^0}$ on the magnetic field, are illustrated
for baryon densities $\rho_0$,  $3\rho_0$ and  $5\rho_0$, 
in figures \ref{mddbar_mag_rhb0}, 
\ref{mddbar_mag_3rhb0} and \ref{mddbar_mag_5rhb0}
respectively. The effects of the isospin asymmetry
on the masses of these open charm mesons are also investigated. 
The in-medium masses of the $D^+$ and $D^0$ mesons are shown 
in panels (a) and (b) of these figures.
The isospin symmetric part ($\sim (\rho_p+\rho_n)$) 
of the first term of the interaction Lagrangian, namely the
Weinberg-Tomozawa term gives a drop in the $D^+$ as well as 
the $D^0$ meson masses
in the hadronic medium, as can be seen from dispersion 
relation (\ref{dispddbar}) and the
self energy for the $D$ meson given by equation
(\ref{selfd}). With isospin asymmetry introduced in the medium, 
the $D^+$ experiences a further drop in the mass, whereas
$D^0$ experiences a positive contribution to the mass,
from the second term of the Weinbrg-Tomozawa term,
thereby giving a mass splitting of $D^+$ and $D^0$ mesons
in the asymmetric nuclear matter. For isospin symmetric
nuclear matter ($\eta$=0), as has already been mentioned,
in the presence of magnetic field, the unequal values 
of the scalar densities for the proton and neutron 
(due to Landau quantization for the proton)
leads to nonzero value of $\delta$. This gives rise
to a splitting in the masses of the $D^+$ and $D^0$
in the presence of magnetic field, arising from the
scalar exchange as well as the range terms, even in symmetric 
nuclear matter. The contributions of all the
terms are observed to show a large drop of the $D^+$ mass
in the asymmetric nuclear medium,
with the drop
being larger for higher asymmetry in the medium.
The $D^+$ meson has a further contribution in the presence
of magnetic field, due to Landau quantization,
which gives a positive shift in its mass, as 
given by equation (\ref{mdpm_landau}).
For the $D^0$ meson, the isospin
asymmetry is observed to give rise to a smaller
drop in the in-medium mass arising from its interaction
with the nucleons and the scalar mesons. The effects of
isospin asymmetry are seen to be quite prominent 
for $D^0$ meson as compared to $D^+$ meson for higher 
values of the densities, 3$\rho_0$ and 5$\rho_0$,
plotted in figures
\ref{mddbar_mag_3rhb0} and \ref{mddbar_mag_5rhb0}.
The effects of anomalous magnetic moment are observed
to give a smaller drop of the masses of the $D^+$ and
$D^0$ mesons, and the difference as compared to when these
effects are not accounted for, is seen to be high 
at high densities.

For $D^+$ mesons, considering AMM effects, the mass obtained from 
dispersion relation, which is further shifted because 
of Landau quantization effects as given by equation 
(\ref{mdpm_landau}), shows a steady increase 
with rising magnetic fields for all densities and 
isospin conditions of the medium. Without AMM effects, 
the increase in the mass is observed to be much less
as compared to when these effects are not taken into account,
and this is more noticeable for higher densities. 
For $\eta$=0.3 and $\rho_B=\rho_0$, for the values of
$eB$ as $m_\pi^2$, $5m_\pi^2$, $8m_\pi^2$, 
the mass of $D^+$  is observed to be 
1783.17 (1781.86), 1807.0 (1802.96) and 1829.2 (1825.4) respectively
with (without) AMM effects.
When the density is increased to $5\rho_0$, the value 
of $D^+$ mass (in MeV) is seen to be 1421.94 (1421), 
1456 (1440) and 1498.8 (1462.6) respectively.

The $D^0$ meson shows only a small modification in its mass 
with respect to change in magnetic field 
at nuclear matter density, as can be seen from figure 
\ref{mddbar_mag_rhb0}. At higher densities, $3\rho_0$ and
$5\rho_0$, plotted in figures \ref{mddbar_mag_3rhb0}
and  \ref{mddbar_mag_5rhb0},
the $D^0$ mass is observed to increase with rise 
in magnetic fields for isospin asymmetric nuclear matter,
when the AMM effects are taken into account.
When AMM effects are not considered, the $D^0$ mass is observed
to remain smaller as compared to when these effects are considered,
and the difference in these masses is seen to reduce
with increase in isospin asymmetry in the medium.  
In the absence of AMM effects, the $D^0$ mass remains constant in case 
of $\eta$ = 0.5, as the magnetic effects for neutrons are only
due to their anomalous magnetic moments. For $\eta$=0.3 
and $\rho_B=\rho_0$, the values of the $D^0$ mass (in MeV)
with (without) AMM effects, 
are found to be 1796.31 (1796.65)
1798.39 (1796.28) and 1798.7 (1795.8) for
$eB$ as $m_\pi^2$, $5m_\pi^2$, $8m_\pi^2$ respectively. 
For $\rho_B=5\rho_0$, these values of $D^0$ mass are modified to
1492.2 (1491.2) 1494.7 (1477.7) and 1506.3 (1472.1) respectively.
For the symmetric as well as asymmetric (with $\eta$=0.3) 
nuclear matter, when the AMM effects are not
included, the mass of the $D^0$ meson is observed to drop
with increase in the magnetic field, and this behaviour is
marginal for $\rho_B=\rho_0$, but quite
appreciable for the higher densities of $3\rho_0$ and $5\rho_0$, 
as can be seen from the figures 
\ref{mddbar_mag_rhb0}, \ref{mddbar_mag_3rhb0}
and \ref{mddbar_mag_5rhb0}. The values of $D^+$ as well as
$D^0$ masses are 
unaffected by magnetic field for $\eta$=0.5, when AMM effects
are not considered, as neutrons have only magnetic field
effects from their AMMs.

The mass of the $D^-$($\bar {D^0}$) meson is observed
to show similar behaviour when the magnetic field is
increased, as the mass of the $D^+$($D^0$) meson,
as can be seen from the figures \ref{mddbar_mag_rhb0}, 
\ref{mddbar_mag_3rhb0} and  \ref{mddbar_mag_5rhb0},
plotted for densities $\rho_0$, $3\rho_0$ and $5\rho_0$
respectively. For $\rho_B=\rho_0$, the mass modification
of the $\bar {D^0}$ remains marginal with rise in the magnetic
field, whereas for higher densities, the value of the $\bar {D^0}$
mass is observed to have appreciable drop when the magnetic field
is increased (but unaffected for $\eta$=0.5), when the AMM effects are not
taken into account. The $D^-$ mass has an increasing trend
with increase in magnetic field, due to the Landau quantization
effects, similar to the $D^+$ mass. The effects of isospin
asymmetry as well as AMM effects are observed to be large 
at higher values of the magnetic fields, and these 
are more pronounced at high densities.
The present study shows that the density effects 
are the most prominent medium effects and the isospin
asymmetry effects as well as AMM effects
are appreciable at high densities. 

\section{summary}
We have studied the medium modifications of the open charm mesons,
namely the $D$ and $\bar D$ mesons, in isospin asymmetric nuclear
matter in the presence of strong magnetic fields. These are
investigated using a chiral effective model, generalized
from SU(3) to SU(4), to obtain the interactions of these
charmed mesons with the light hadrons. The medium modifications
of the masses of these mesons in the asymmetric nuclear matter
arise due to their interactions with the nucleons and the 
scalar mesons ($\sigma$, $\zeta$ and $\delta$). The number density
and scalar density of the proton have contributions
from the Landau energy levels in the presence of the magnetic field.
The effects of the anomalous magnetic moments of the proton and
neutron are also investigated in the present work. 
The effects of the isospin asymmetry as well as the 
anomalous magnetic moments of the nucleons are observed
to be quite prominent, especially at high densities.
The mass modifications of the $D^+$ and $D^0$ mesons
of the $D$ meson doublet, as well as of $D^-$ and
$\bar {D^0}$ of the $\bar D$ doublet,
are observed to be quite different in the isospin
asymmetric medium at high densities.
The effects of magnetic field on the $D$ and $\bar D$
meson masses are observed to be much less as compared to
the effects from density, in the present work. 
The density effects on the in-medium masses 
of these open mesons are thus the dominant medium effects, 
which should be observed in the $D^+/D^0$ and
$D^-/{\bar {D^0}}$ ratios, as well as in the
partial decay widths of the charmonium states
to $D^+D^-$ and $D^0 \bar {D^0}$ pairs, in asymmetric 
heavy ion collisions planned at the Compressed
Baryonic Matter (CBM) experiment at FAIR at the 
future facility at GSI.

\acknowledgements
The authors thank Hiranmaya Mishra for fruitful discussions. 
One of the authors (AM) is grateful to the Institut 
f\"ur Theoretische Physik for warm hospitality and acknowledges financial 
support from Alexander von Humboldt Stiftung when this work was initiated. 
Amal Jahan CS acknowledges the support towards this work from 
Department of Science and Technology, Govt of India, 
via INSPIRE fellowship
(Ref. No. DST/INSPIRE/03/2016/003555).

\end{document}